\documentclass[3p,twocolumn]{elsarticle}

\usepackage{amssymb}
\usepackage{amsmath}
\usepackage{lineno}
\usepackage{xcolor}
\journal{Chaos Solitons and Fractals}

\begin{document}


\begin{frontmatter}

\title{Group-size dependent synergy in heterogeneous populations}

\author[label1]{Hsuan-Wei Lee}
\cortext[mail1]{Correspondence should be addressed to: szolnoki.attila@ek-cer.hu or hwwaynelee@gate.sinica.edu.tw}
\author[label1]{Colin Cleveland}
\author[label2]{and Attila Szolnoki}

\address[label1]{Institute of Sociology, Academia Sinica, Taiwan}
\address[label2]{Institute of Technical Physics and Materials Science, Centre for Energy Research, P.O. Box 49, H-1525 Budapest, Hungary}

\begin{abstract}
When people collaborate, they expect more in return than a simple sum of their efforts. This observation is at the heart of the so-called public goods game, where the participants' contributions are multiplied by an $r$ synergy factor before they are distributed among group members. However, a larger group could be more effective, which can be described by a larger synergy factor. To elaborate on the possible consequences, in this study, we introduce a model where the population has different sizes of groups, and the applied synergy factor depends on the size of the group. We examine different options when the increment of $r$ is linear, slow, or sudden, but in all cases, the cooperation level is higher than that in a population where the homogeneous $r$ factor is used. In the latter case, smaller groups perform better; however, this behavior is reversed when synergy increases for larger groups. Hence, the entire community benefits because larger groups are rewarded better. Notably, a similar qualitative behavior can be observed for other heterogeneous topologies, including scale-free interaction graphs.
\end{abstract}

\begin{keyword}
Public goods game \sep cooperation \sep heterogeneous groups
\end{keyword}

\end{frontmatter}

\section{Introduction}
\label{intro}

The conflict between individual and collective interests can be captured by several social dilemma games, including the prisoner's dilemma, snowdrift, and stag-hunt games. In the simplest case, players can choose between two options: to cooperate or defect, and the highest individual income can be reached if a player defects. However, if all rational players follow this choice, the cooperation terminates in a state that is frequently referred to as the ``tragedy of the commons'' \cite{hardin_g_s68}. Unfortunately, such situations are witnessed too frequently, with examples including global warming, overexploitation, environmental problems, and tax evasion. Therefore, it is a vital and practical intellectual task to identify the mechanisms and conditions that can help avoid this scenario and support the evolution of cooperation among selfish individuals.

The aforementioned games represent situations in which the players' interactions can be described by two-point functions. However, the original dilemma can also be detected when a group of players makes simultaneous decisions that determine their payoff values. Such multi-point interactions are faithfully described by the public goods game, where group members can still decide whether they cooperate and contribute to a common pool or defect and refuse this sacrifice \cite{perc_jrsi13,he_jl_pla22,xiao_sl_epjb22,zheng_jj_pa22,liu_jz_epjb21,zhong_xw_pa21}. Importantly, the accumulated contributions are multiplied by an $r>1$ synergy factor, which represents synergy among group members. In other words, the product of two or more players could be larger than the simple sum of their efforts, because they could work more effectively in a group. It is also a crucial element of the public goods game, in which the enhanced individual productivity is distributed evenly among group members, regardless of whether they contributed to the common pool. Therefore, the original dilemma remains harsh if the synergy factor is too small and cooperators only have a chance to survive if $r$ exceeds the group size.

Several extensions of the traditional public goods game have been suggested in the last decade, and scientists have identified interesting aspects that could be essential for driving evolution toward a higher level of cooperation \cite{szolnoki_epl12,quan_j_csf21,yang_lh_csf21,szolnoki_pre10b,wang_xj_csf22,szolnoki_csf22,flores_jtb21,quan_j_jsm20,szolnoki_epl16}. The first milestone was to recognize that a spatially structured population, in which players have fixed and limited interactions with others, could help introduce network reciprocity among cooperators \cite{nowak_n92b,szabo_pr07,perc_pr17}. Thus, these players can reach competitive payoff values against the defectors. It was also revealed that heterogeneity among participants could also be a cooperator supporting condition because it facilitates the spatial coordination of strategies \cite{perc_pre08,santos_n08}. The latter, however, can reveal the collective benefits of the cooperative strategy. Speaking of heterogeneity, it could also be beneficial if a cooperative player treats the neighboring groups differently and distributes their potential contributions unequally among the groups involved. A cooperator may prefer neighbors who organize a larger group venture \cite{wang_hc_pa18,cao_xb_pa10}; however, the focal player can also reward groups that provide a greater reward earlier \cite{quan_j_pa21,zhang_hf_pa12,vukov_jtb11}. Monitoring the strategies of neighboring players and rewarding players who have a history of cooperation can be useful for cooperators  \cite{yang_hx_pa19,yuan_wj_pone14,quan_j_jsm22}. However, incrasing cooperation by supporting the most successful neighbor without knowing their strategy is less straightforward \cite{szolnoki_amc20,lee_hw_pa21}. To disseminate the enlarged contributions, we may expect a certain cooperation level, which is a type of threshold. This condition could also be useful for motivating players to be more cooperative \cite{wang_j_pre09,szolnoki_pre10,chen_q_csf16,szolnoki_pre12,jaegher_srep20,du_jm_fp18,deutchman_ehb22,wang_sx_cnsns19,du_cp_amc23}. But alternative approaches, like utilizing reputation or applying hetrogeneous resource allocation could also result in higher cooperation \cite{han_csf22,gao_amc21,meloni_rsos17,wei_x_epjb21}.

From the viewpoint of our present work it is worth stressing that the potential impact of group size of public goods game on the cooperation level was already studied by several earlier papers \cite{han_aa17,szolnoki_pre11,segbroeck_prl12,han_if15}. Notably, in all the aforementioned models, the synergy factor is generally assumed to be a uniform parameter. However, this assumption can be easily criticized because the collaboration of a small or large groups should not necessarily be equally effective. For instance, it is our everyday life experience that a larger company is more effective; hence, it can easily crowd out smaller competitors from a market. Furthermore, cooperation is not only organized in small communities but naturally extends to larger groups, including firms, towns, countries, or larger regions.

Therefore, exploring the consequences of assuming that the synergy factor, as the key parameter of the public goods game, depends on the group size is reasonable. Based on our argument, we suppose that the larger the group, the greater the synergy factor. Evidently, this effect should have a limit; therefore, we assume a limited group size above which all groups have the same maximum synergy factor. Another interesting question could be how the synergy factor changes when the group size is increased. From this aspect, we introduce different options where this parameter grows linearly or nonlinearly when the group size is increased.

\section{Improving synergy in larger community}
\label{def}
 
\begin{figure}
	\centering
	\includegraphics[width=7.0cm]{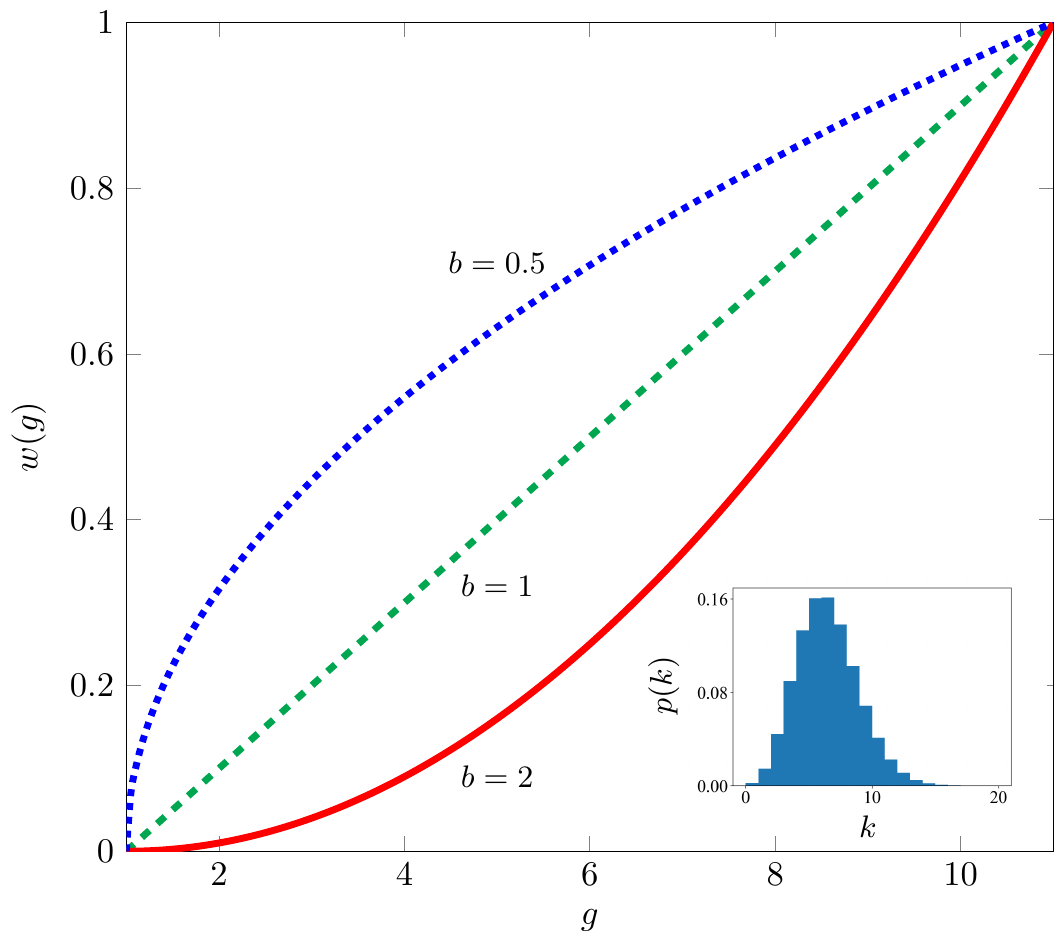}\\
	\caption{Dependence of the weight factor of synergy on the group size for different $b$ values. For small $b$, a smaller group could be effective, while a large $b$ represents situations where mutual efforts can only be effective for large groups. The inset shows the degree distribution in a heterogeneous population modeled by a random graph used in this work. Notably, the group-size distributions follow a similar Poisson distribution.}\label{weight}
\end{figure}

To explore how group-size-dependent enhancement factors change evolutionary outcomes, we assume a heterogeneous population in which players are involved in games with different group sizes. The simplest topology that fulfills this condition is a random graph, where the average degree of the nodes is $\langle k \rangle$. In the following, we assume that $\langle k \rangle =6$; hence, every player is involved in seven games on average: in a group where she is the focal player, and in $\langle k \rangle = 6$ other games organized by the neighbors. The degree distribution follows a Poisson distribution, as illustrated in the inset of Fig.~\ref{weight}. Here, we can see that only a few groups have more than 11 members; therefore, we assume that the group size limit is 11. Beyond this size, all the groups enjoy the maximum synergy factor without a discount. However, below this threshold size, smaller groups have a reduced enhancement factor. Technically, this can be achieved by introducing a $w(g) \le 1$ weight factor, which is then multiplied by $r$ if the $g$ group size is below the $G=11$ limit. 

A further question is the actual shape of the $w(g)$ function. There are three conceptually different options. According to the simplest choice, $w(g)$ increases linearly with group size. In an alternative case, an early increase in $g$ results in a significant improvement in the weight factor, and hence, the synergy. In the third $b=2$ case, the maximal impact can only be reached in the vicinity of a large group-size limit. All three cases can be handled using a unified formula for the weight factor as follows:
\begin{equation}
w(g) = \left(\frac{g-1}{G-1}\right)^b\,,
\end{equation}
where $G$ is the threshold size of the large groups where this effect saturates. The shape of this function for different values of the exponent $b$ is illustrated in Fig.~\ref{weight}. This figure shows that for $b=2$, small groups have very limited efficiency, and the appropriate impact can only be reached for large groups. However, when $b$ is small, smaller groups have almost the same effectiveness as larger groups.

One may claim that if we reduce the synergy factor for certain groups, we manipulate the general cooperation level artificially because a smaller enhancement factor supports the defector strategy. Therefore, to make our results comparable, we introduced an effective synergy factor, which is calculated as follows:
\begin{equation}
	r_{eff} = \sum_{g=2} p(g) w(g) r \,,
\label{eff}
\end{equation}
where $p(g)$ denotes the probability of finding a group of size $g$ in a population. By using the same $r_{eff}$ for different $b$ parameter values, we can properly compare the resulting cooperation levels with the traditional model, where a uniform synergy factor is used for all groups.

In the following, we use a standard protocol to monitor the strategy evolution. First, we generated a random graph with $N$ nodes, where the average degree was $\langle k \rangle =6$. The typical system size is $N=5000$; however, we also checked different values to avoid finite-size problems. Initially, every node was randomly assigned as a cooperator or defector. During the elementary step, we select player $i$ and neighboring player $j$. If their strategies are different, we calculate their payoff values. The payoff of player $i$, who is involved in $k_i+1$ games, is calculated as
\begin{equation}
	\Pi_i = \sum_{m=0}^{k_i}\frac{w(g_m) r n_{C_m}}{g_m} - c_i\,.
\end{equation}
Here, $m=0$ denotes the group where player $i$ is the focal player, while $m=1 \dots k_i$ represents groups where the neighbors are focal players; hence, player $i$ is also involved. Furthermore, $g_m$ denotes the size of group $m$, and the number of cooperators in this group is $n_{C_m}$. The contribution of player $i$ to a common pool is 1 if $i$ is a cooperator and 0 otherwise.

According to the standard protocol, player $i$ adopts the strategy of player $j$ with a probability
\begin{equation}
\Gamma (j \to i) = \frac{1}{1+\exp(\Pi_i-\Pi_j)/K}\,,
\end{equation}
where $K$ denotes the noise parameter of the decision. Without the loss of generality, we apply $K=0.1$, which makes our results comparable to previous model calculations made on spatial public goods game. In this way we can monitor the consequence of size-dependent synergy directly. At this point, it is also worth noting that the noise parameter or its inverse, which is frequently called selection strength, may play a decisive role in human experiments \cite{rand_pnas13,zisis_srep15}. A full Monte Carlo step ($MCS$) involves repeating the above-described elementary step $N$ times; hence, on average, every player has a chance to update their strategy. To reach the stationary state, the typical relaxation time is varied between $10^4$ and $10^5$ $MCS$s before we measure the stationary fraction of the cooperators. For good statistical analysis, the entire procedure was repeated 500 times.

\section{Results}

\begin{figure}[h!]
	\centering
	\includegraphics[width=7.0cm]{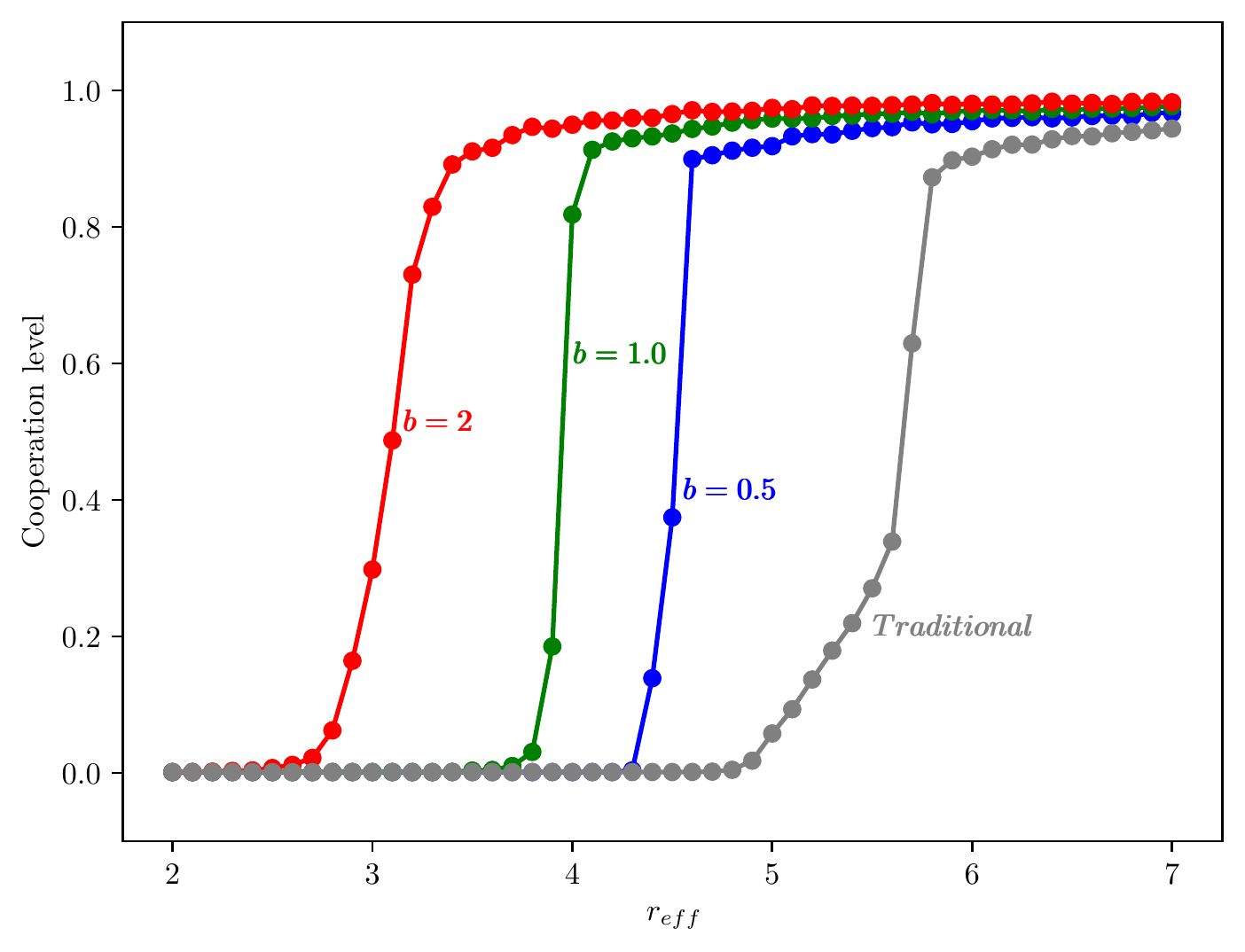}\\
	\caption{Dependence of the general cooperation level on the effective synergy factor. The curve, marked by ``traditional", depicts the case when all groups have an identical synergy factor independent of their sizes. The largest improvement can be reached for $b=2$, where only large groups are rewarded by a relatively higher synergy factor. The results are averaged over 500 independent runs.}\label{ER_all}
\end{figure}

Our key observations are summarized in Fig.~\ref{ER_all}, where we plotted the general cooperation level, which is the average fraction of cooperative players, for different cases. Here, the symbol sizes are comparable to the variance of averages, while curves are used as eye guides. ``Traditional'' denotes the case when all groups apply the same synergy factor; hence, $w(g)=1$ for all groups. We stress that the cooperation levels are plotted as a function of $r_{eff}$, as defined by Eq. ~\eqref{eff}. This simply means that we must apply a higher $r$ value in Eq. ~\eqref{eff} for $b=2$ than for $b=0.5$ to reach the same average cooperator-supporting environment. Otherwise, a comparison would make no sense. Evidently, $r_{eff}=r$ in the traditional case.

It is better for the entire community if synergy depends on the group size. Furthermore, the best improvement can be achieved for the $b=2$ case, in which only large groups are awarded the extra reward of cooperation. Here, the critical $r_{eff}$ value, where cooperators can survive, is almost half the value observed for the traditional model, where every group works with the same efficiency. 

\begin{figure}
	\centering
	\includegraphics[width=7.0cm]{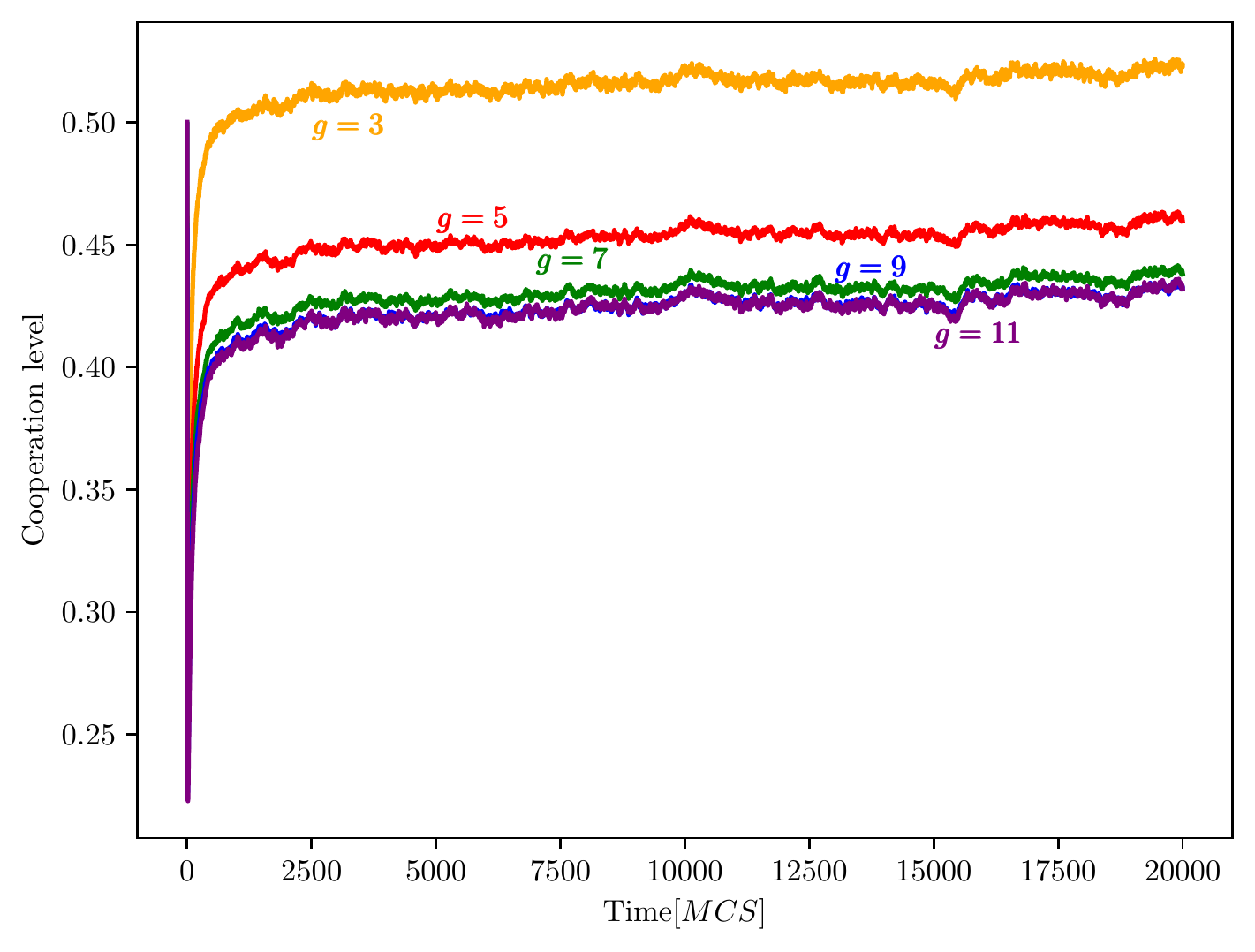}\\
	\caption{Typical time evolution of cooperation levels in groups with different sizes in the traditional model at $R_{eff}=R=5.11$. The size of groups are indicated. Smaller groups reach higher cooperation levels. }\label{trad_evol}
\end{figure}

To identify the possible mechanism that explains this improvement, we monitored how the cooperation level evolved locally in specific groups. More precisely, we measured the cooperation level in groups of different sizes. For comparisons, as shown in Fig.~\ref{trad_evol}, we first present the evolution of these values in the traditional model when evolution is launched from a random initial state. In agreement with previous observations, the cooperation level first decays very quickly, but after the remaining cooperators start building protective clusters in the sea of defectors. Notably, cooperation increases more intensively in the smallest groups and always ensures better conditions for cooperators. This is a generally valid result that has been previously confirmed from different perspectives \cite{wang_z_pre12b,ohtsuki_n06,wang_z_srep12}. 
 
\begin{figure}
	\centering
	\includegraphics[width=7.0cm]{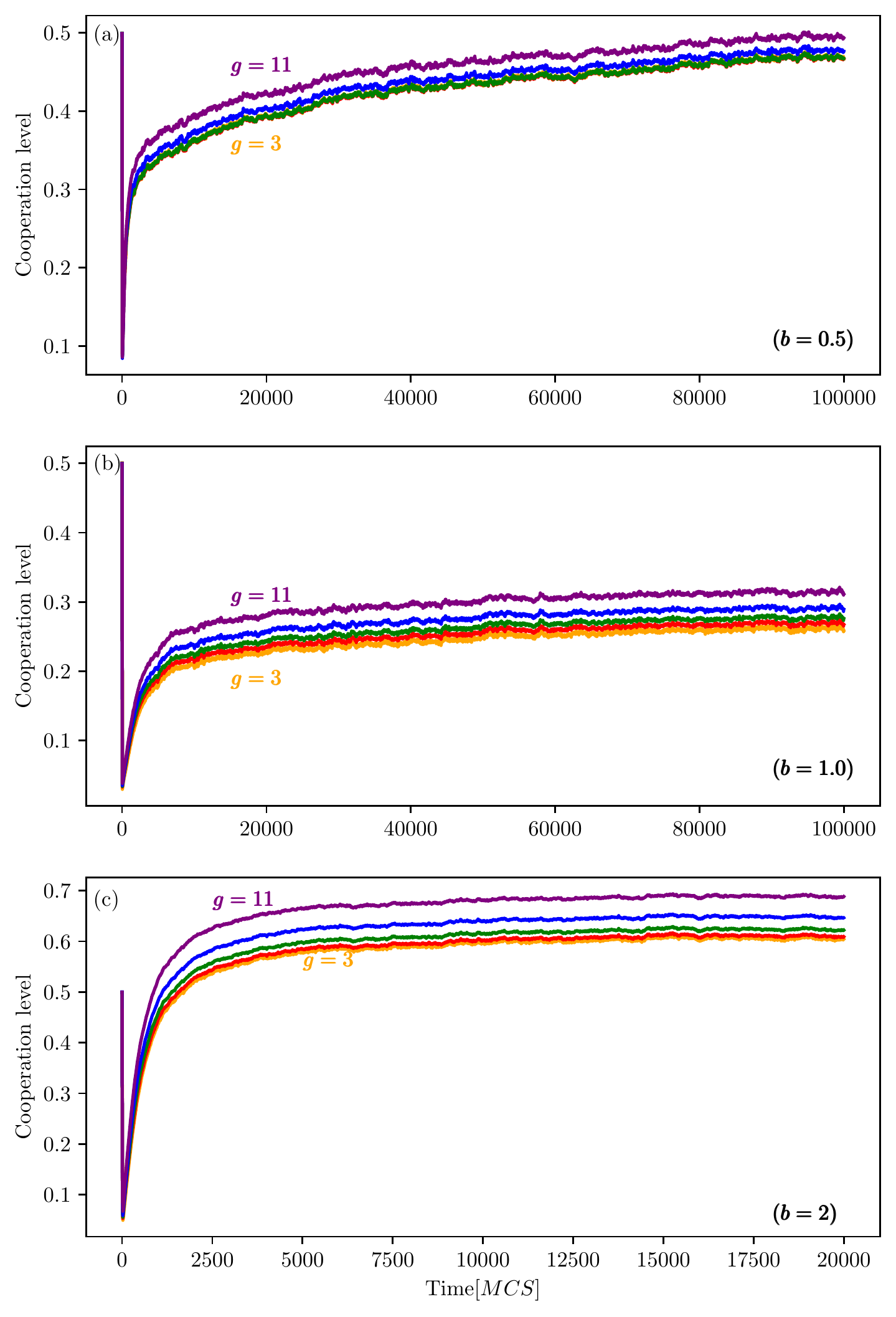}\\
	\caption{ Representative time evolution of cooperation levels in groups with different sizes in the model where the synergy factor is group-size dependent. The $b$ values are indicated, while $R_{eff}=4.497$, 3.93, and 3.51 from top to bottom, respectively. These values were selected to provide comparable general cooperation levels for all cases. The group size for the highest and lowest cooperation levels are shown on the related curves. For $b=2$, shown in bottom panel, larger groups perform noticeably better than smaller groups.}\label{ER_evol}
\end{figure}

However, what if larger groups work more efficiently? The results are shown in Fig.~\ref{ER_evol}. In the early stage, we detect similar system behavior as previously reported because defectors enjoy a random environment; hence, the cooperation level decays in every type of group. However, after this short initial period, cooperation emerges more easily in larger groups owing to the larger $w(g)$ weight factors. Larger groups performed better, which is beneficial to the entire community. In other words, the rank of cooperation among groups of different sizes reverses and becomes the opposite of the traditional case. This effect can be observed for all $b$ values, as shown in Fig.~\ref{ER_evol} but is more pronounced for $b=2$, as shown in panel~(c).

As a conclusion, group-size dependent synergy could be useful because larger groups are less vulnerable to the invasion of defectors. In this way, in the areas, where maintaining cooperation would be the biggest challenge in a uniform system, we can reach significant improvement. However, smaller groups can still maintain a certain cooperation level even if the actual synergy factor is smaller. In summary, the entire population gains more if larger groups are supported at the expense of smaller communities.

\begin{figure}
 	\centering
 	\includegraphics[width=7.0cm]{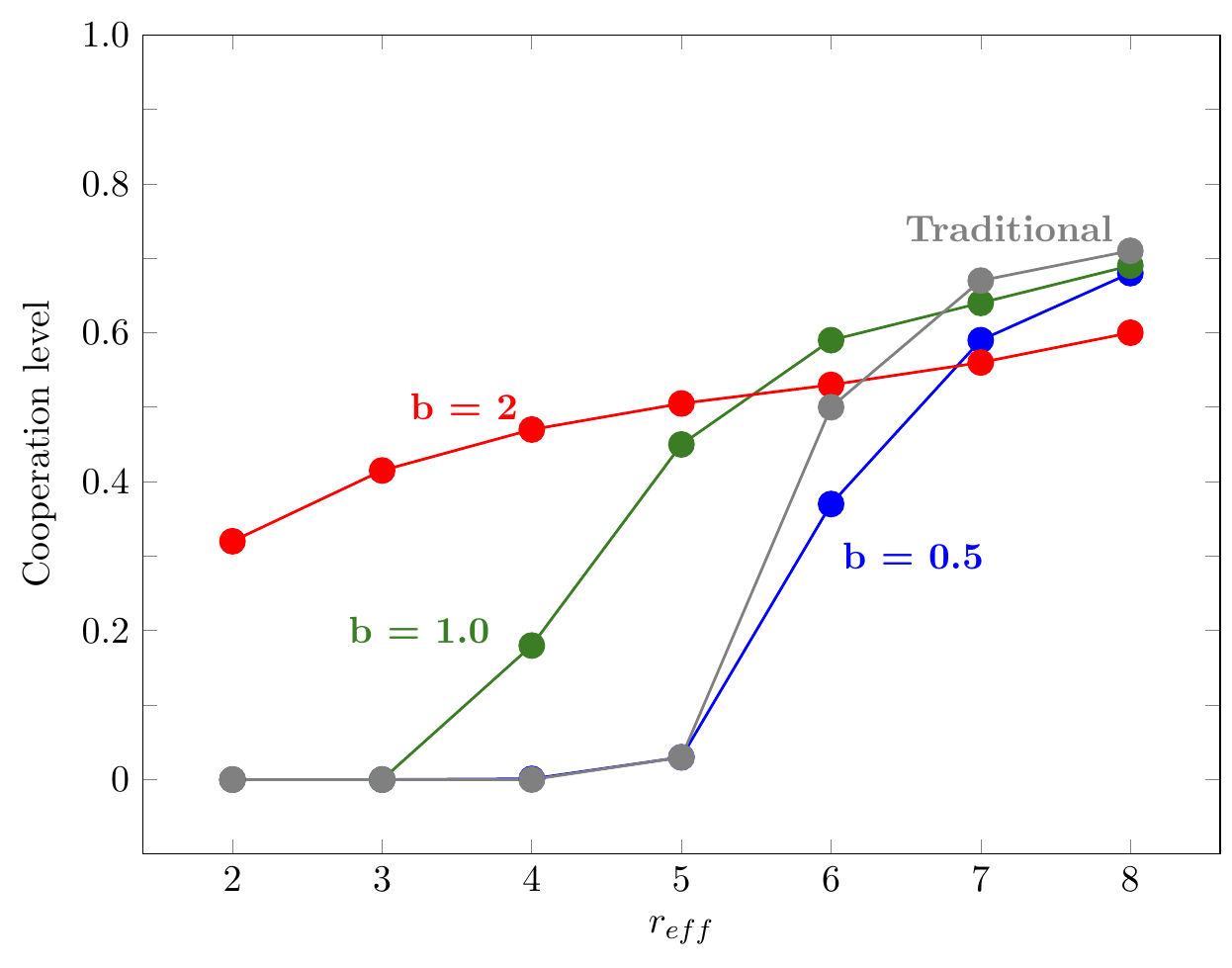}\\
 	\caption{The general cooperation level depending on the effective synergy factor obtained for scale-free interaction graphs where we used the same $\langle k \rangle =6$ average degree applied when networks were generated. Larger symbols mark larger fluctuations in the averages. The results are averaged over 800 independent runs.}\label{sf}
 \end{figure}

If we apply other types of heterogeneous interaction topologies, conceptually similar behaviors can be observed. For instance, we can use a scale-free topology where degree distribution is even more heterogeneous than for Erd{\H o}s-R{\'e}nyi random graph. In this case, our observations are summarized in Fig.~\ref{sf} where we used the same $\langle k \rangle =6$ average degree as previously. This extra heterogeneity results in significantly larger fluctuations in the averages, which are signed by larger symbol sizes in this plot. But the condition when large groups work more efficiently provides a significant improvement when the dilemma is harsh, which means in the low $r_{eff}$ region. We must note that the system behavior is less surprising compared to the random graph topology. In the former case, hubs with a large degree can collect significantly larger payoffs. This advantage eventually helps them homogenize the strategy distribution in their neighborhoods. While it is also valid for defector hubs, in the long term, it is beneficial for cooperator hubs because it helps augment network reciprocity \cite{szabo_pr07,santos_n08}. A reasonable consequence is that larger groups behave more cooperatively than smaller groups in the traditional system, where an equally strong social dilemma is used everywhere \cite{szolnoki_pa08}. Therefore, in this case to ``support'' larger groups with larger synergy will not change the original rank. Indeed, some improvements were still observed, as it is illustrated in Fig.~\ref{sf}. For instance, cooperators die out at $r=r_{eff}=4$ in the traditional model, but the cooperation level is approximately 0.2 for $b=1$ and is approximately 0.5 for $b=2$; however, the entire effect is not as spectacular as for random graphs. Last, we also note that the ``failure'' of $b=0.5$ case also supports our argument, because in this case, the larger groups become less favored compared to the large $b$ cases, hence the original enhanced network reciprocity is weakened.

\begin{figure}
 	\centering
 	\includegraphics[width=7.0cm]{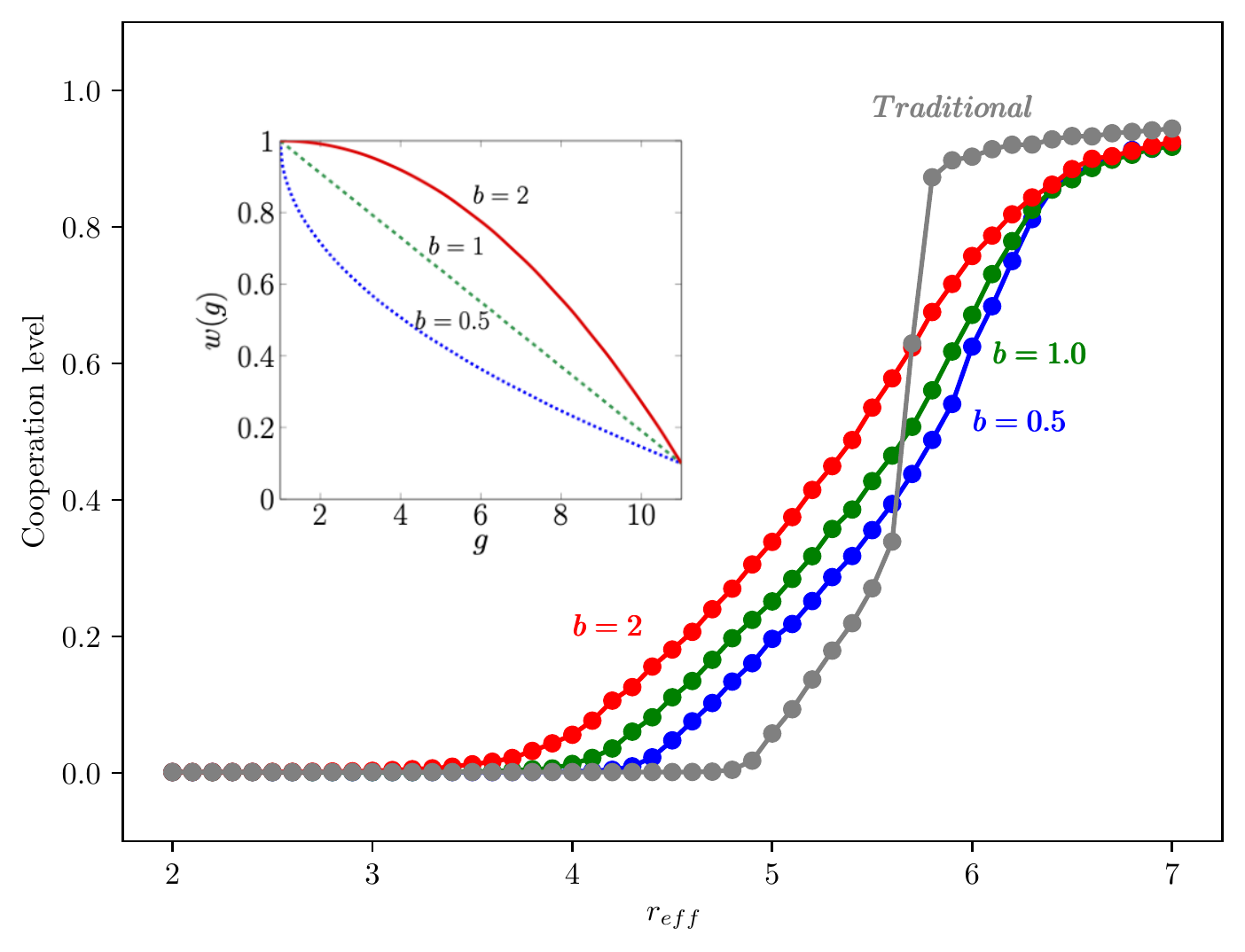}\\
 	\caption{The general cooperation level depending on the effective synergy factor for when the weight factor is a decaying function of group size. The $w(g)$ functions for different $b$ values are shown in the inset. In this ``reversed" case, group-size dependent synergy has no clear advantage over the uniform model. The results are averaged over 500 independent runs.}\label{reverse}
 \end{figure}

One may claim that group-size-dependent synergy is not necessarily a growing function. In a small community, players are controlled more easily; hence, defection is revealed immediately. This effect can be captured by a relatively high synergy factor for smaller groups. In the framework of our model, this can be achieved using a decreasing function for the weight factor. The form of this function is shown in the inset of Fig.~\ref{reverse}, where we assume that groups larger than $g \ge G=11$ can only use a minimal, $w=0.1$, weight factor. However, smaller groups are given higher weight factors. As previously mentioned, we can distinguish three conceptually different cases of how weight function decays.

The main panel presents the resulting general cooperation levels. This comparison suggests that cooperation is not unambiguously supported when synergy decays with group size. Indeed, it shows a slight increment for small $r_{eff}$ values; however, for higher $r_{eff}$ values, the uniform model performs better. Furthermore, the best results can be obtained for $b=2$ when the advantage of smaller groups is less pronounced. These observations implicitly confirm that it is more beneficial if larger groups are more effective than smaller groups. More precisely, if we support smaller groups at the expense of larger formations, we render the system more vulnerable to defection. In this case, larger groups cannot be competitive even for larger $r_{eff}$ and the entire community will be punished. On the other hand, additional support does not have a significant effect on small groups that are ready for cooperation at a modest $r$ value. More precisely, it causes some improvement for small $r_{eff}$ values, where the cooperators would not survive in the traditional model. However, this improvement is the highest for $b=2$, when not only the smallest groups are rewarded by a significant enhancement factor but the ``support'' is more balanced among groups with different sizes.

\section{Conclusions}

Public goods game, as a paradigm of social dilemmas, has already been examined by several research papers from different perspectives \cite{yu_fy_csf22,wang_jw_pla22,shen_y_pla22,kang_hw_pla21,deng_ys_pa21,allen_pre18}. Many of these studies assumed that players form equally large groups to create social goods \cite{yang_hx_pa19,pei_zh_njp17,quan_j_c19,fu_mj_pa19,xu_zj_c19,liu_rr_amc19}. Evidently, it is more realistic to consider a population comprising groups of different sizes. In the latter case, however, assuming that all types of groups are equally effective is the simplest approach. We can easily explain why larger (smaller) groups are more effective, which can be characterized by a larger synergy factor. Motivated by this idea, in this study, we introduced a model in which groups use a synergy factor that is strongly related to group size. 

First, we assumed that larger groups enjoy higher synergy factors and found that the heterogeneity of smaller communities can generate a significantly higher general cooperation level. This effect is highest when only those groups can enjoy a significantly high enhancement factor whose size is close to or above a threshold value. Importantly, for a proper comparison, we must use an effective enhancement factor, which is the weighted average of all applied values in the population.

To reveal the possible reasons for this improvement, we monitored how the cooperation level evolves in groups of different sizes. Notably, smaller groups perform better in the traditional model, where all group sizes have the same $r$ enhancement factor. When larger groups use higher $r$ values, the rank of groups is reversed, which results in a higher general cooperation level. Therefore, it can be inferred that higher support from larger groups can restore the vulnerability of these communes, which is beneficial to everyone.  

If we apply the inverse protocol and support smaller groups better, we cannot observe a clear improvement. More precisely, there is an increase in cooperation for smaller values of effective $r$; however, for higher enhancement factors, the group-size-dependent $r$ model performs worse than the traditional model. In this case, the additional support of small groups cannot increase significantly the cooperation level in these small communities. However, the suppression of larger groups via smaller $r$ values worsens the situation in the latter groups, which is already critical in the traditional setup.

In summary, our model underlines the possible positive role of larger groups; hence, we can identify an independent mechanism that supports the evolution of cooperation. Nevertheless, we should not forget the advantage of a small community, in which individual actions are more visible. Therefore, defects can be detected and punished more easily. However, a larger group that functions more effectively can resolve the original problem without assuming additional incentives, such as punishment or reward \cite{perc_srep15,lv_amc22,li_my_csf22,ohdaira_srep22,wang_sx_amc22,cong_r_srep17,yang_hx_epl20,szolnoki_pre15,wang_sx_pla21, cheng_f_amc20,gao_sp_pre20}. Therefore, we can conclude that the entire community gains more if larger groups are more efficient than smaller groups. This finding confirms our daily life observations and explains why human societies spontaneously organize into larger structures. This biased efficiency is beneficial to the entire system.

\vspace{0.5cm}

This research was supported by the Ministry of Science and Technology of the Republic of China (Taiwan) under Grant No. 109-2410-H-001-006-MY3 and the National Research, Development and Innovation Office (NKFIH) under Grant No. K142948.

\bibliographystyle{elsarticle-num-names}

\end{document}